\documentclass[useAMS,usenatbib]{mn2e}
\usepackage{amsmath,graphicx,psfig}

\title[Anomalous SZ Contribution to  WMAP]{Anomalous SZ Contribution
to 3 Year WMAP Data} \author[R.M. Bielby and T. Shanks]
{R.M. Bielby\thanks{E-mail:r.m.bielby@durham.ac.uk} and T. Shanks\\
Department of Physics, Durham University, South Road, Durham, DH1 3LE,
UK\\}
\begin{document}
\date{\today}
\pagerange{\pageref{firstpage}--\pageref{lastpage}} \pubyear{2006}
\maketitle
\label{firstpage}
\begin{abstract}
We first  show that the new WMAP 3 year data confirm the detection by
\citet{myers04} of an extended SZ signal centred on 606 Abell (ACO)
clusters with richness class, $R\geq2$. Our results also show SZ
decrements around APM and 2MASS groups at increased significance than
previously detected. We then follow the approach of \citet{lieu06a}
and compare the stacked WMAP results for the decrement in 31 clusters
with ROSAT X-ray profiles where Lieu et al found on average less SZ
decrement in the WMAP 1 year data than predicted. We confirm that in
the 3 year data these same clusters again show less SZ decrement than
the X-ray data predicts. We then analysed the WMAP results for the 38
X-ray clusters with OVRO/BIMA measured SZ decrements as presented by
\citet{bon06}. We again find that the average decrement is measured to
be significantly less (5.5$\sigma$) than predicted by the Chandra
X-ray data. Thus while we confirm the original detection of an
extended SZ effect by \citet{myers04}, these X-ray comparisons may now
suggest that the central SZ amplitudes detected by WMAP may actually
be lower than expected. One possible explanation is that there is
contamination of the WMAP SZ signal by radio sources in the clusters
but we argue that this appears implausible. We then consider the
possibility that the SZ decrement has been lensed away by foreground
galaxy groups.  Such a model predicts that the SZ decrement should
depend on cluster redshift. A reduction in the SZ decrement with
redshift is suggested from the ACO cluster sample and also from
comparing the samples of \citet{lieu06a} and \citet{bon06}. However,
the mass power-spectrum would require a far higher amplitude than
currently expected if lensing was to explain the SZ deficit in high
redshift clusters.
\end{abstract}
\section{Introduction}

\citet{myers04} made a cross-correlation analysis between galaxy
cluster catalogues and the WMAP first year data \citep{wh}. They saw a
statistical decrement near groups and clusters as detected by APM and
also in  more nearby groups and clusters as detected by 2MASS but the
strongest signal was seen in the ACO rich cluster catalogue. There the
decrement was approximately  what was expected from predictions based
on X-ray observations of the Coma cluster which is itself a richness
class 2 cluster. However, the profile appeared to be more extended
than expected from simple fits to these typical cluster X-ray
data. The extent of the SZ effect, possibly to $\theta\approx1$
degree, led \citet{myers04}  to speculate whether the SZ effect could
contaminate the measurement of the acoustic peaks, although the
difference between the SZ and primordial CMB spectral indices may
constrain such a possibility at least for the first peak
\citep{huff04}. We now return to this topic with the first aim to see
if the extended SZ effect reproduces in the 3-year WMAP data.

Meanwhile, \citet{lieu06a} analysed the WMAP first year data now
focusing only on 31 clusters with ROSAT X-ray data. They made basic
predictions for the SZ decrement in each cluster and found that they
over-predicted the SZ decrement. One possibility was that discrete
radio sources in the clusters were diluting the decrements but this
was argued against by \citet{lieu06a}. However, \citet{lieu06b}
suggested an alternative mechanism based on synchrotron radiation from
cosmic ray electrons moving in the cluster magnetic field forming a
diffuse cluster radio source which again may dilute the SZ
effect. This model was also aimed at explaining the soft X-ray
excesses detected in some clusters  via inverse Compton scattering of
the CMB by the same cosmic ray electrons in the cluster
(e.g. \citealt{neva03} and references therein).

Here we shall check the result of \citet{lieu06a} using our
cross-correlation methodology and the full WMAP 3-year data. In the
first instance, we shall take the X-ray models of \citet{lieu06a}
which follow the simple $\beta$ model prescription described in
Section 3 below. We shall also look at a new sample of clusters with
excellent Chandra X-ray data \citep{bon06}. Again we shall simply take
their models convolved for the WMAP PSF in the appropriate band and
compare to the averaged SZ decrement seen in the WMAP3 data.

\section{Data}
\subsection{WMAP Third Year Data}
In this paper we use the raw CMB temperature maps provided in the WMAP
3 year data release \citep{hin06}. These consist of temperature data
from the five frequency bands and the internal linear combination
(ILC) map (Table~\ref{wmapbands}). In order to remove contamination
from our own galaxy, we make use of the Kp0 foreground mask made
available with the other WMAP data products and have applied this to
all our maps prior to cross-correlation. Finally, the data is used
here in the HEALPix format of equal area data elements, characterised
by N$_{side}$=512, which gives an element width of $\approx7'$.

\begin{table}
 \centering
  \caption{Properties of the WMAP frequency bands.}
  \label{wmapbands}
  \begin{tabular}{@{}lrr@{}}
  \hline Band & Frequency & FWHM \\ \hline W  & 94GHz & $12.'6$ \\ V &
  61GHz & $19.'8$ \\ Q  & 41GHz & $29.'4$ \\ Ka & 33GHz & $37.'2$ \\ K
  & 23GHz & $49.'2$ \\ \hline
  \end{tabular}
\end{table}

\subsection{Cluster Data}
\subsubsection{ACO}
The ACO catalogue \citep{aco} lists clusters with 30 or more members,
given the requirements that all members are within 2 magnitudes of the
third brightest cluster member, whilst also lying within a 1.5
h$^{-1}$ Mpc radius. A richness class, R, is applied to the individual
clusters based on a  scale of 0 $\le$ R $\le$ 5. The catalogue covers
both hemispheres and here we trim these samples such that we take
clusters of only R $\ge$ 2 and galactic latitudes of $\mid$b$\mid$
$\ge$ 40$^{\circ}$.

\subsubsection{APM}
We shall also use galaxy group and cluster catalogues derived from the
APM Galaxy Survey of \citet{mad} which covers the whole area with
$\delta<-2.5\,$deg and $b<-40\,$deg.  These were identified using the
same `friends-of-friends' algorithm as \citet{adm} and references
therein.  Circles around each APM galaxy with $B<20.5$ are `grown'
until the over-density, $\sigma$, falls to $\sigma=8$ and those
galaxies whose circles overlap are called groups.  The APM galaxy
surface density is $N\approx750\,$deg$^{-2}$ at $B<20.5$. Minimum
memberships, $m$, of $m\geq 7$ and $m\geq 15$ were used. The sky
density of groups and clusters is 3.5$\,$deg$^{-2}$ at $m\geq7$ and
0.35$\,$deg$^{-2}$ at $m\geq 15$. We assume an average redshift of
$z=0.1$ for both APM samples.

\subsubsection{2MASS}
The third cluster catalogue is derived from the final data release of
the 2MASS Extended Source Catalogue (XSC) \citep{jar} to a limit of
$K_s\leq 13.7$. $K$-selected galaxy samples are dominated by
early-type galaxies which are the most common galaxy-type found in
rich galaxy clusters. Therefore the 2MASS survey provides an excellent
tracer of the high density parts of the Universe out to $z<0.15$ and
so provides a further test for the existence of the SZ effect. Using
the above 2-D friends-of-friends algorithm, \citet{myers04} detected
500 groups and clusters with $m\ge35$ members at the density contrast
$\sigma=8$ in the $|b|\ge10\,$deg area. The 2MASS groups have  average
redshift, $z\approx0.06$.

\subsubsection{ROSAT X-ray  cluster sample}
The 31 clusters published by \citet{bon02} were originally selected as
a sample of X-ray bright clusters suitable for observing X-ray surface
brightness profiles. These profiles were obtained with the ROSAT PSPC
instrument and estimates of the gas temperature, density and
distribution were made by fitting a $\beta$ profile model to the data
(see section 3 below). The X-ray data for these 31 clusters were
previously used by \citet{lieu06a} to construct predictive models of
the SZ profile of each cluster. Redshifts for the clusters range from
z$\sim$0.02 (Coma) up to z$\sim$0.3 (Abell 2744), whilst the sample
lies in the galactic latitude range of $\mid b\mid \ge 25^{\circ}$.

\subsubsection{Chandra X-ray cluster sample}
We further analyse the 38 clusters discussed by \citet{bon06}. These
clusters have been observed at 30GHz by OVRO and BIMA (see
\citet{bon06} and references therein) to detect the SZ decrements and
have also been observed by Chandra to provide the X-ray data needed to
estimate the value of H$_0$. The interferometric  radio observations
have a resolution of $\approx1'$ and the X-ray observations from the
Chandra ACIS-I camera have a resolution of $\approx1''$. Redshifts for
these clusters are in the range $0.18<z<0.8$, a higher range than for
the ROSAT sample. \citet{bon06} fitted both hydrostatic equilibrium
and isothermal $\beta$ models to the X-ray data and made predictions
for the SZ decrements.

\begin{figure*}
 \includegraphics[width=160.mm]{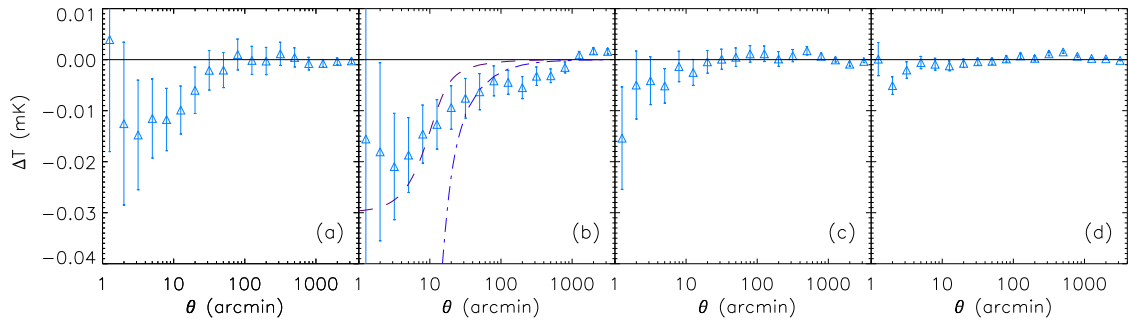}
 \caption{Cross-correlation results between the WMAP 3-year W-band
 temperature data and the four cluster datasets: (a) 2MASS, (b) ACO,
 (c) APM m$\geq$15, and (d) APM m$\geq$7. The dashed and dot-dashed
 lines in (b) show SZ models with $\Delta T = 0.083K$ and $\Delta T =
 0.49K$ respectively, both with $\theta_c = 1.`5$ and $\beta = 0.75$
 and convolved with the WMAP beam-width. The latter model is intended
 to be representative of the Coma cluster, scaled  to redshift
 $z=0.15$. The former is the  ACO model fitted by \citet{myers04} in
 their analysis of the WMAP 1st year results.}
 \label{wmap-all}
\end{figure*}

\begin{figure}
 \includegraphics[width=75.mm]{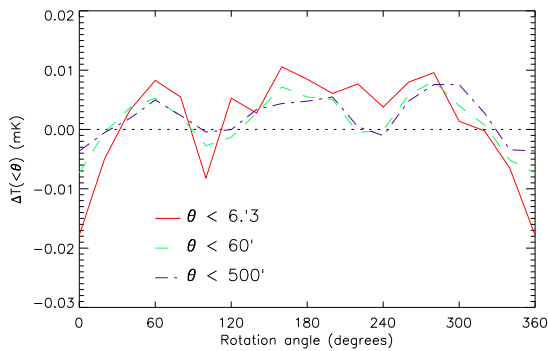}
 \caption{The cross-correlation of the ACO catalogue is shown after
 increments in galactic longitude of 20$\deg$ in the Abell cluster
 positions. The mean $\Delta T$ is shown for WMAP pixels within
 $6.'3$, $60'$ and $500'$ of cluster centres, where the significance
 at each angular limit is $3.2\sigma$, $2.0\sigma$ and $1.2\sigma$
 respectively.}
 \label{rotation}
 \end{figure}

\section{SZ X-ray modelling}
The SZ effect is generally modelled using  X-ray gas profiles,
densities and temperatures. The X-ray data is most simply modelled by
fitting a $\beta$ model to the X-ray intensity profile:
\begin{equation}
S_X = S_{X0} \left( 1 + \frac{\theta^2}{\theta_c^2} \right)
^{(1-6\beta)/2}
\label{xbeta}
\end{equation}
where $S_{X0}$ is the central X-ray surface brightness and $\theta_c$
is the angular core radius. On the isothermal assumption, the
temperature decrement, $\Delta T_{SZ}$, as a function of the angular
distance from the cluster-centre, $\theta$, is then given by:
\begin{equation}
\Delta T_{SZ}(\theta)~=~\Delta
T_{SZ}(0)\left[1+\left(\frac{\theta}{\theta_c}\right)^2\right]^{-\frac{3\beta}{2}+\frac{1}{2}}
\label{sztheta}
\end{equation}
Then the magnitude of the central temperature shift, $\Delta
T_{SZ}(0)$, is given by:
\begin{equation}
\frac{\Delta T_{SZ}(0)}{T_{CMB}}~=~\frac{kT_e}{m_ec^2}\sigma_{Th} \int
\mathrm{d}l \quad n_e \left[\frac{x(e^x+1)}{e^x-1}-4\right]
\label{dt0}
\end{equation}
\noindent where $x=h\nu/kT_e$ , $\sigma_{Th}$ is the Thomson
cross-section and $n_e$, $T_e$ are the gas density and temperature
derived from the X-ray data.

\citet{lieu06a} use the cluster sample of \citet{bon02} and fit ROSAT
PSPC cluster X-ray profiles. They assume isothermal gas distributions
with $T_e$ taken from \citet{bon02}. \citet{bon06} use both a
hydrostatic equilibrium model, allowing a double power-law
$\beta$-model to allow for variations in the number density with
radius, and an isothermal $\beta$-model. With the hydrostatic model,
they allow the gas temperature to vary with radius and a CDM component
as well as gas to contribute to the cluster potential. We shall simply
assume the isothermal models of \citet{lieu06a} and \citet{bon06} and
convolve the predicted SZ profile with the appropriate WMAP beam
profile, modelled as a Gaussian with the FWHM beam-widths shown in
Table~\ref{wmapbands}.

\begin{figure*}
 \includegraphics[width=140.mm]{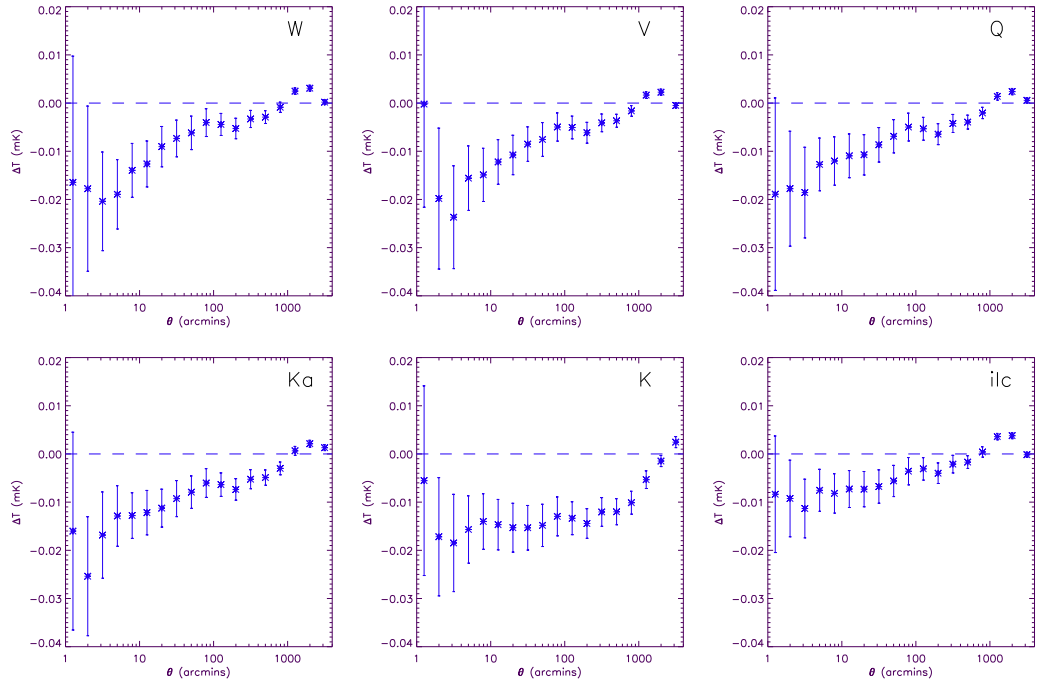}
 \caption{Cross-correlation results between 606 ACO rich galaxy
 clusters ($R\ge2$, $|b|>40$deg) and the WMAP 3-year maps in  5
 band-passes ($+$ILC) as indicated.}
 \label{WMAP-ACO}
\end{figure*}

\section{Cross-correlation analysis}
We focus our analysis on the 94GHz W band from WMAP, looking for
correlations characteristic of the SZ effect in this, the highest
resolution band. We perform a cross-correlation analysis as described
in \citet{myers04}, calculating the mean temperature
decrement/increment as a function of angular separation from galaxy
clusters in the above datasets. Our cross-correlation takes the form:
\begin{equation}
   \Delta T_c(\theta) = \sum_{\substack{i}} \frac{\Delta
   T_i(\theta)-\overline{\Delta T}}{n_i(\theta)}
\end{equation}
Where $\Delta T_i(\theta)$ is the WMAP temperature in an element $i$
at an angular separation $\theta$ from a cluster centre and $n_i$ is
the number of elements at that separation. $\overline{\Delta T}$ is
the mean WMAP temperature decrement across the entire region used in
the analysis. For the 3-year W-band data, $\overline{\Delta
T}\sim10^{-3}$mK.

Errors on our results are estimated using repeated Monte Carlo
realisations of the cluster data. As each catalogue (i.e. 2MASS, ACO,
APM) will each be highly clustered, it is important to incorporate
this clustering into the realisations. Thus for each realisation we
begin by creating a set of random positions with 5$\times$ the number
of clusters as in the parent catalogue. We then calculate a clustering
amplitude for each individual cluster in this random sample and
attribute a weighting based on the assigned clustering
amplitude. Clusters are then selected or discarded based on this
weighting, until the realisation has the required number of selected
clusters (i.e. that of the parent catalogue). As a final check, the
auto-correlation of the realisation is measured. Through comparison
with the auto-correlation of the original catalogue, the realisation
is either accepted or discarded. This process is repeated until we
have 100 acceptable clustered mock catalogues for each parent
catalogue. The cross-correlation is then calculated between the WMAP
data and the 100 mock catalogues and the standard deviation is taken
as the 1$\sigma$ error on our results.

In addition to this we also perform a rotational analysis to provide
an alternative estimate of the errors. In this case we perform the
cross-correlation between the cluster positions and the WMAP data. We
then shift the cluster positions by 20$\deg$ in galactic longitude and
recalculate the cross-correlation. We repeat this until we have
rotated through a full 360$\deg$. A S/N is then calculated from the
results of this rotational analysis.

\section{Results}
\subsection{Optical/IR Cluster Samples}
The results for the cross-correlation between the four large cluster
datasets (APM m$\geq$7, APM m$\geq$15, ACO R$\ge$2 and 2MASS clusters)
and the WMAP W-band data are shown in Fig.~\ref{wmap-all}. A decrement
is immediately evident on small scales within $\theta <30'$ of cluster
centres in all four data sets. Looking in detail first at the ACO
results, the WMAP3 cross-correlation strongly confirms the results of
\citet{myers04} from WMAP 1st year data. Here, we find a decrement of
$-0.021\pm0.007 mK$ at $\theta < 6'.3$ and $-0.010\pm0.004 mK$ at
$\theta < 60'$ for the W-band data (quoted accuracies are from the
Monte-Carlo analysis). Basically, the ACO decrement has remained the
same and the improved statistics at small angles has increased the
S/N. In addition to the Monte-Carlo analysis, we also checked our ACO
results using the rotational analysis described by \citet{myers04} and
find the significance  of the decrement at $6.'3$, $60'$ and $500'$ to
be $3.2\sigma$, 2.0$\sigma$ and 1.2$\sigma$ (see Fig. \ref{rotation}).
As before, there appears to be some form of extended signal out to
angles of $\sim100'$. Following \citet{myers04}, we also produce the
correlations with the four other WMAP bands, plus the ILC map. The
results of this are shown in Fig.~\ref{WMAP-ACO}. Again, good
agreement is seen between these updated results and the original 1st
year data results. Despite the increasingly poor resolution of the
bands, the decrement is observed in the V, Q and Ka bands, whilst even
the ILC map and the Ka band map show a decrement.

Improvements in the small scale statistics are also observed in the
2MASS and APM results while  the magnitudes of the decrements remain
unchanged. However, the APM group ($m\ge7$) SZ detections remain
marginal  even at small scales.

\subsection{ROSAT X-ray bright cluster sample}
\begin{figure}
 \centering \includegraphics[width=40.mm]{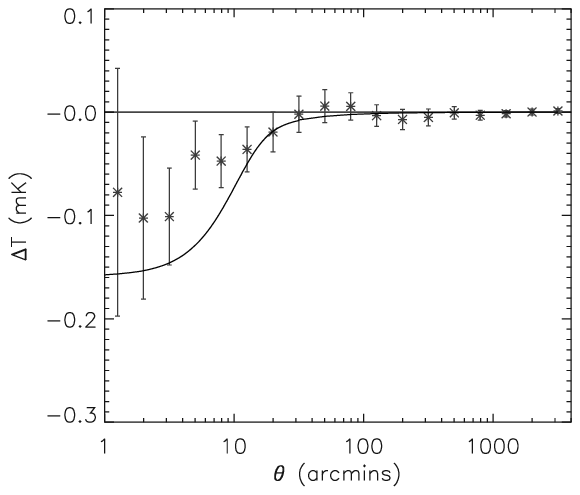}
 \includegraphics[width=40.mm]{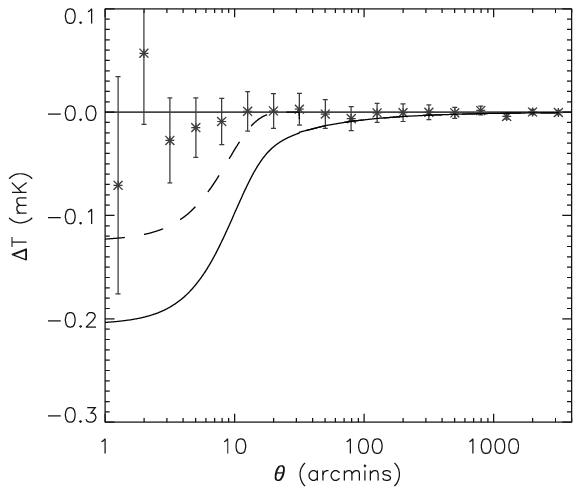}
 \caption{Average $\Delta T$ (from WMAP W-band data) plots for 30
 clusters from the ROSAT sample (left) and 39 clusters from the
 Chandra sample (right). In both figures, the points show our
 cross-correlation results, whilst the curves show average SZ models
 (based on the parameters taken from \citealt{lieu06a} and
 \citealt{bon06}) convolved with a Gaussian representing the WMAP beam
 profile. For the Chandra sample, we plot the full isothermal model
 (solid line) and the same model limited to $\theta<2'$ (dashed line).}
 \label{lieubon}
\end{figure}

We next consider the X-ray bright clusters of \citet{bon02}. Analysis
of this dataset with respect to the first year WMAP results has
already been performed by \citet{lieu06a}. Their main conclusion was
that the SZ decrement in the WMAP1 data around the locations of these
clusters has a lower magnitude than they would expect from their
predictions based on the original X-ray observations of
\citet{bon02}. In Fig.~\ref{lieubon} (left panel) we show our
cross-correlation between the 31 clusters used by \citet{lieu06a} and
the WMAP year 3 data in the W-band (crosses). We also present the
average model prediction based on the \citet{bon02} data (solid line).
This has been convolved with a Gaussian beam profile of $\sigma=6.'3$.
We see the same general effect as seen by \citet{lieu06a}, that the SZ
effect is somewhat smaller than predicted by the data. However, the
significance of rejection is only $\approx2\sigma$. Similar results
are seen in the other WMAP bands.

\begin{figure}
 \includegraphics[width=40mm]{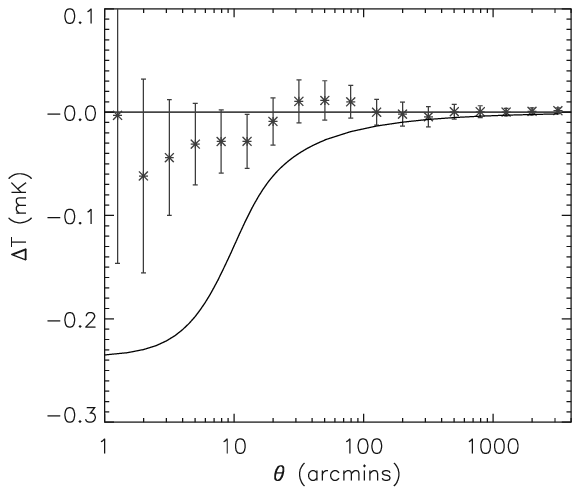}
 \includegraphics[width=40mm]{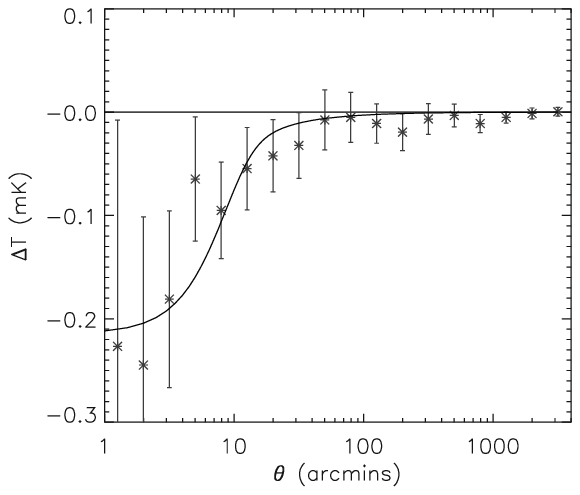}
 \caption{Average $\Delta T$ (from WMAP W-band data) correlations for
 the ROSAT X-ray clusters split by redshift: 21 clusters at $z<0.1$
 (left) and 9 at $z>0.1$ (right).}
 \label{lieuzb}
\end{figure}

We next split the \citet{lieu06a} clusters by redshift as shown in
Fig.~\ref{lieuzb}. Again our results are given by the crosses, whilst
the solid lines show the average model SZ profiles. The model for
clusters at $z<0.1$ is rejected by $4.2\sigma$ at $\theta<6.'3$,
whilst at $0.1<z<0.3$ the rejection drops to $1.6\sigma$ at
$\theta<6.'3$. We have also performed this analysis with a latitude
split at $|b|=40^{\circ}$ and find some degeneracy between latitude
and redshift as many of the low redshift clusters are also at low
latitude. However, in either case we do not regard the difference
between the results in Fig.~\ref{lieuzb} as particularly statistically
significant.

\subsection{Chandra X-ray bright cluster sample}

\begin{figure}
\centering \includegraphics[width=40.mm]{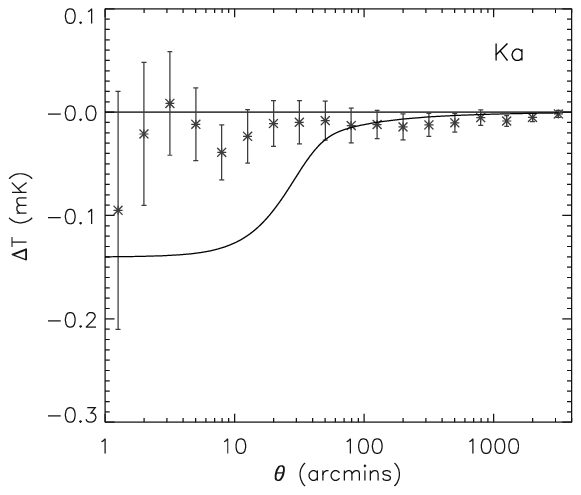}
 \includegraphics[width=40.mm]{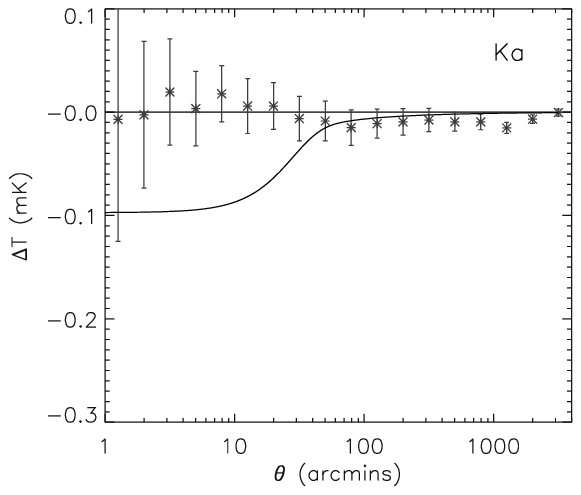}\\
 \includegraphics[width=40.mm]{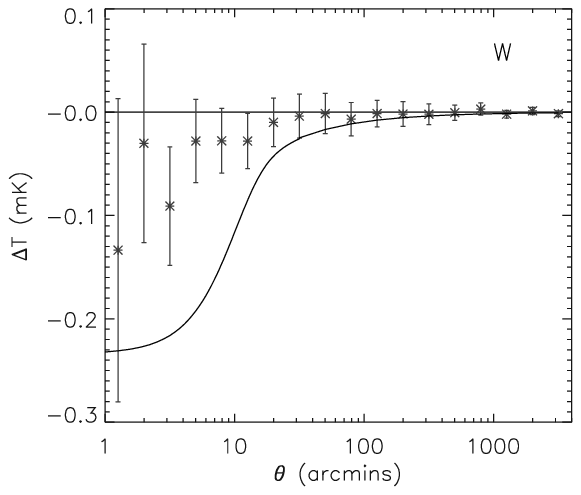}
 \includegraphics[width=40.mm]{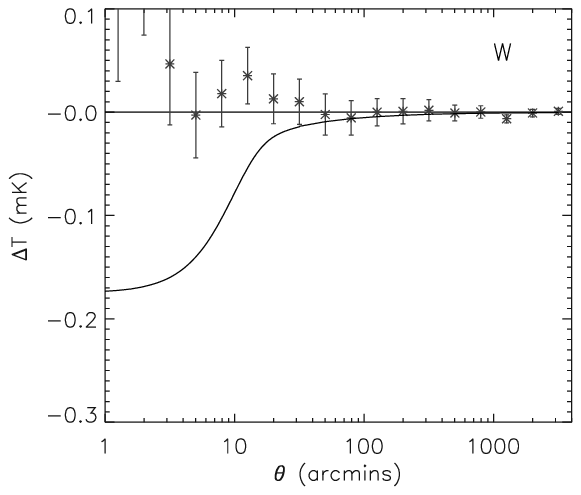}
 \caption{Top: Cross-correlations of WMAP Ka-band $\Delta$T data with
 20 clusters at $z<0.3$ (left) and 19 at $z>0.3$ (right) from the
 Chandra cluster sample. Bottom: The same for WMAP W-band data. The
 solid lines show the $\beta$ models of \citet{bon06} convolved with
 the WMAP profiles.}
\label{bonkw}
\end{figure}

We next analysed the SZ decrements for the 38 clusters of
\citet{bon06}, using the WMAP3 W band results. In Fig.~\ref{lieubon}
(right panel) we compare the cross-correlation results with an average
model constructed from the individual isothermal models given in Table
5 of \citealt{bon06} (solid line) and again find that the SZ effect is
now quite severely over-predicted by the models, with a rejection
significance of $5.5\sigma$. We again looked for a dependence on
redshift and found slight evidence for a greater SZ signal at $z<0.3$
compared to to $z>0.3$ (Fig.~\ref{bonkw}).

Given that \citet{bon06} only fit the Chandra data for $\theta<2'$,
there is the possibility that this model may not apply at the large
angles covered by the WMAP data. We therefore also show in
Fig.~\ref{lieubon} the SZ model truncated at $\theta=2'$ before being
convolved with the W-band beam (dashed line). The significance of
rejection in this case is reduced to $2.5\sigma$. We note that this is
a strict lower limit to this significance limit as it assumes no SZ
contribution beyond $2'$.

\begin{figure}
 \centering \includegraphics[width=75.mm]{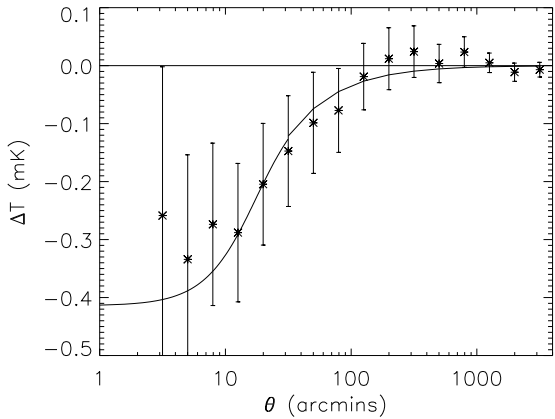}
 \caption{Binned $\Delta T$ data from the WMAP year-3 W-band data
 around the Coma cluster. The solid line shows the model predicted
 from X-ray data (taken from \citealt{lieu06a}) convolved with the
 W-band beam profile.}
 \label{coma}
\end{figure}

\begin{figure}
  \centering \includegraphics[width=40.mm]{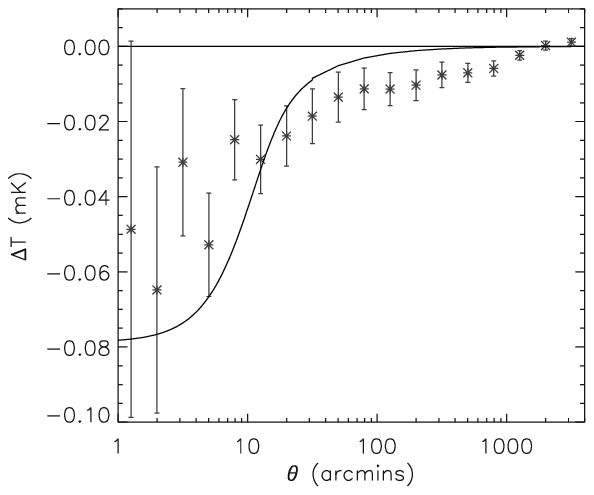}
  \includegraphics[width=40.mm]{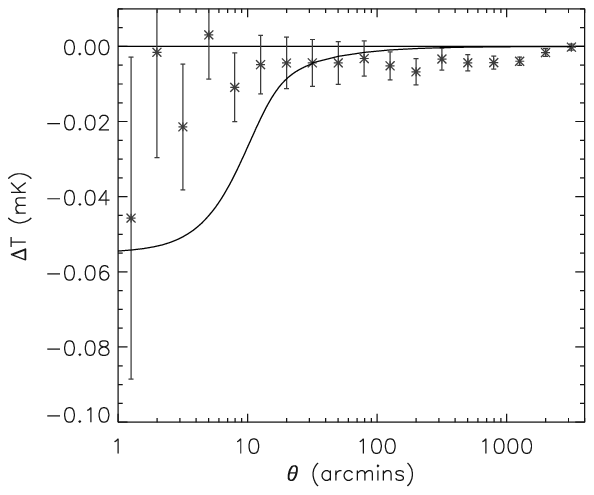}
  \caption{Average $\Delta T$ (from WMAP W-band data) plots for the
  data from the Abell cluster catalogue with 172 clusters at $z<0.15$
  (left) and 235 at $z>0.15$ (right). Only clusters with
  $\mid$b$\mid\ge40\deg$ are included here. Overlaid in both cases is
  a model with $\Delta T(0) = -0.16mK$, $\beta = 0.7$ and $\theta_c =
  9.'8$ scaled to the mean redshift of the samples: $z = 0.1$ and $z =
  0.2$. In both cases the model is convolved with the W-band beam
  profile. This model gives a reasonable fit to the data at $z<0.15$,
  but significantly overestimates the $z>0.15$ data.}
  \label{acoz}
\end{figure}

Although within the Chandra sample there is little evidence of
redshift dependence, the fits of SZ models to the WMAP data do appear
to deteriorate as we move from the average redshift, $z\approx0.1$ of
the ROSAT sample to $z\approx0.3$ of the Chandra sample
(Fig.~\ref{lieubon}). We have also noted that at the lowest redshift
the WMAP SZ effect is clearly detected at about the predicted
amplitude in the Coma cluster (Fig.~\ref{coma}). We therefore returned
to the ACO dataset and identified 407 $R\ge2$, $|b|>40$deg  clusters
with measured redshifts. Splitting these at z=0.15 (Fig.~\ref{acoz}),
we see that there is some evidence confirming that clusters at higher
redshift have observed SZ decrements that are significantly smaller
than at lower redshift. Although the X-ray properties for the majority
of these clusters are unknown, we have fitted the same average model,
scaling $\theta_c$ to the appropriate average redshift before
convolving with the WMAP beam. The fit appears significantly worse for
the higher redshift clusters, with a rejection confidence at
$\theta<6.'3$ of $\approx1\sigma$ for $z < 0.15$ and $\approx4\sigma$
for $z > 0.15$. We tentatively conclude that there may be a redshift
dependence of the SZ effect in the sense that higher redshift clusters
show a smaller than expected effect.

\section{Discussion}

The reduced SZ decrements in the WMAP3 data towards the ROSAT cluster
sample and the almost lack of detection of the SZ effect  in terms of
the \citet{bon06} clusters is paradoxical. The most obvious
explanation is that the WMAP data is contaminated by unresolved
cluster radio sources within the WMAP beam. However, the contamination
from synchrotron radio point sources varies with frequency as
$T_{\nu}\propto\nu^{\alpha-2}$ (where $\alpha\approx0.7$), whilst the
discrepancy in the WMAP3 data for the \citet{bon06} cluster sample is
as large at Ka (33GHz) as at W (94GHz) (see Fig.~\ref{bonkw}).

Further to this issue, we note that a survey of radio sources in the
Chandra clusters has been performed by \citet{cob07}. They see a
population of radio sources with a mean flux of $\approx
6.2$mJy/cluster at 30GHz. Given a spectral index of such sources of
$\alpha\approx0.7$ (where $S_{\nu}\propto \nu^{-\alpha}$) this gives a
flux of $\approx3$mJy/cluster at the W-band frequency of
90GHz. Following \citet{lieu06a}, the equivalent flux required to
cause the lack of SZ effect observed in the Chandra clusters can be
determined from the Rayleigh-Jeans flux multiplied by the solid angle:
\begin{equation}
S_{SZ} = \frac{2\pi k\Delta T \nu^2}{c^2}\frac{\theta^2}{4}
\end{equation}
Taking $\Delta T = 0.1$mK, $\nu = 90$GHz and $\theta = 10'$, we obtain
a flux of $S_{SZ} = 170$mJy. Even taking a value for the spectral
index of $\alpha=0$ for the radio sources (e.g. \citealt{benn03}), the
flux required is over an order of magnitude greater than the observed
discrete radio source fluxes from \citet{cob07}. In addition,
\citet{linmohr} make estimates of the contamination from radio point
sources and for cluster masses typical of the Chandra sample
($M_{200}\sim 10^{15}M_{\sun}$, \citealt{reibohr}), they suggest that
up to only 10\% of these clusters may be lost due to point source
radio contamination (see their Fig. 15). Although this assumes that
there will be no increase in source contamination with the WMAP beam
area, we note that the counts of \citet{cob07} are usually lower than
predicted by \citet[][ Fig.~13]{linmohr} and these effects may cancel.

Currently we have no explanation for the strong SZ decrements detected
by the interferometric experiments  as opposed to the lack of
detections by WMAP. At higher resolution it may be more possible to
detect the SZ against the noise caused by the primordial CMB
fluctuations but our error analysis should take care of such
statistical effects and the average model is rejected at the
$5.5\sigma$ level by the WMAP data. A high  value of
$H_0\approx100$kms$^{-1}$Mpc$^{-1}$ for the SZ X-ray model might help
explain the ROSAT cluster results but an even higher value would be
required to explain the Chandra cluster results.

As noted above there may also be evidence that the SZ decrement is too
low in the ACO-WMAP1 cross-correlation of \citet{myers04}, as
confirmed by the ACO-WMAP3 cross-correlation in Fig.~\ref{wmap-all}.
\citet{myers04} noted that the decrement that fitted the ACO $R\ge2$
clusters with $\beta=0.75$ was only $\Delta T(0)=0.083$mK compared to
the 0.5mK predicted for the $R=2$ Coma cluster. The WMAP3 data
confirms that $\Delta T(0)=0.5$mK is needed to fit the observed Coma
SZ decrement (see Fig.~\ref{coma}). In Fig.~\ref{wmap-all} the SZ
models for these two values of the  decrement are compared to the
WMAP3 W band data for the $R\ge2$ cluster sample. Both models assume
$\beta=0.75$. We see that while the data is well fitted at
$\theta<10'$ by the $\Delta T(0)=0.083$mK model, the $\Delta
T(0)=0.5$mK at least begins to improve the fit at larger scales. One
possibility is that, as well as detecting an extended SZ component to
the ACO data, we may actually be detecting a lower central SZ
amplitude than expected from the X-ray data.

\citet{lieu06a} discussed other possible explanations for the
unexpectedly small SZ decrements detected in the ROSAT sample. For
example, \citet{lieu06b} have discussed whether a diffuse cluster
synchrotron source could explain the reduced SZ decrement. The main
problem here is that non-thermal electrons would not give a good fit
to the X-ray data which are usually well fitted by thermal
bremsstrahlung, although \citet{lieu06b} also noted that the soft
X-ray excess seen in the central regions of some clusters may be
indicative of a significant embedded non-thermal X-ray component there.

\citet{fos03} have discussed whether ISW effect could mask the SZ
effect but the ISW effect is at 0.5$\mu$K and seems too small to mask
the SZ effect which in the X-ray clusters  can be 10$\times$ higher.

There is also the possibility that the SZ decrement has been
overestimated by the X-ray modelling. Certainly the Chandra predicted
decrements  for the 5 clusters in common with the ROSAT sample (A665,
A1413, A1689, A1914, A2218) are on average  $\approx80$\% larger than
the predicted decrements from the ROSAT data. Most of this  difference
arises from A2218 where the ROSAT data imply $\Delta T(0)=-0.27$mK
(corrected to 30GHz) and the Chandra data imply $\Delta T(0)=-0.87$mK,
a factor 3.2 different. But since Chandra has higher spatial
resolution, it is expected to probe the central core of a cluster more
accurately and so the Chandra X-ray models might be expected to be
more robust than those from the ROSAT data.

While this paper was in preparation, \citet{afs06} have also used
X-ray data of 193 Abell Clusters to search for  the SZ decrement from
WMAP3 data (see also \citealt{afs05}). These authors made a
significant detection and also suggested that the size of SZ decrement
implied that the cluster hot gas fraction was $32\pm10$\% lower than
the baryon fraction in the standard cosmological model. They also
suggested that their WMAP results were consistent with the
interferometric SZ results for the sample of 38 Chandra clusters
analysed above. Note that the approach of \citet{afs06} is different
from that used here in that the X-ray data is mainly used to define a
template to detect  SZ decrements and then the SZ data and the X-ray
temperature data alone are used to establish the gas densities. This
route therefore avoids comparing the SZ results with X-ray gas density
models on the grounds that the latter depend on assumptions such as
that of hydrostatic equilibrium. These authors also do not consider
the possibility that the cluster  SZ decrements may evolve with
redshift.

Finally, if we assume that the WMAP SZ decrements are reliable, even
in the case of the 38 Chandra clusters where the unexplained
discrepancy persists with the OVRO/BIMA results of \citet{bon06},  we
might speculate whether  a lower than expected  SZ decrement in the
higher redshift clusters could  be caused by foreground lensing. The
indication from WMAP that the higher redshift clusters may have
reduced SZ decrements is consistent with the idea that gravitational
lensing is having a significant effect on the detection of the SZ
effect. Therefore it may be that the groups and clusters out to
redshifts in the range $0.2<z<0.8$ in the foreground of the targeted
Chandra clusters  are lensing the cluster centres and smoothing the
decrement away. Using CMBFAST we have constructed the lensing
smoothing function for  CMB scattering at $z=0.3$ and find that on the
size of the $\approx10$ arcmin WMAP beam, the smoothing function is
reduced by about a factor of $\approx10$ compared to the  case where
the surface of last scattering is at $z=1100$. At $z=0.7$, the factor
is $\approx5$. Therefore for the standard model this would make the
effect negligible because at z=0.3,
$\epsilon=\sigma/\theta\approx0.004$ and at $z=0.7$,
$\epsilon=\sigma/\theta\approx0.008$. Only if the mass power spectrum
is significantly higher than that for the standard model can this
explanation apply. One such case is the high mass power spectrum
advocated by Shanks (2006) as a route to modify the first acoustic
peak in the CMB. Such a spectrum is motivated by the evidence from QSO
lensing that the galaxy distribution is strongly anti-biased
($b\approx0.1$) at least  on $0.1-1$h$^{-1}$Mpc scales with respect to
the mass \citep{adm, myers05,gm07}. However, the balance of other
evidence may still argue against such a high amplitude for the mass
power-spectrum.

Lensing would clearly also affect the X-ray cluster profiles as well
as the SZ decrements. Although these are expected to be smoother than
the SZ decrements, it might be expected that the profiles of lower
redshift clusters are on average steeper than the profiles of higher
redshift clusters. It remains to be seen whether this prediction of
the lensing hypothesis can be decoupled from evolution of the cluster
gas component.  In any case, the flatness of the X-ray profiles
towards the centres of many clusters may make this prediction more
difficult to test.

\section{Conclusions}

We have confirmed the extended appearance of the SZ decrement in WMAP
3-year data around ACO $R\ge2$ clusters out to $\theta\approx30'$,
first shown by \citet{myers04} using WMAP 1-year data. Further to
this, we have confirmed the detection of the SZ decrement in the
3-year data around clusters identified in both the APM survey and
2MASS, showing an increase in detection significance compared to the
1-year data analysis.

We have also confirmed the result of \citet{lieu06a} that the SZ
decrement is somewhat lower than expected on standard model
assumptions and ROSAT X-ray profiles for a sample of 31 clusters from
\citet{bon02}. We have further shown that even smaller X-ray
decrements are seen in the higher-redshift sample of 38 clusters of
\citet{bon06} that has Chandra X-ray data.  The reason for the
observational discrepancy between the WMAP data and the BIMA/OVRO data
of \citet{bon06} is not clear. We do not believe that  discrete or
diffuse cluster radio sources nor  the ISW effect  is likely to
explain the discrepancy. Dividing the ACO clusters into high and low
redshift samples also indicates that the deficit in the SZ decrement
may increase at higher redshift.

In the light of the above results from our WMAP SZ analysis, we have
discussed the possibility that the extended SZ signal detected for ACO
and 2MASS  clusters may actually be indicating a lack of SZ signal in
the centres of clusters rather than an excess at the edges.

On the assumption that the WMAP SZ results are correct, one
explanation we have considered is that lensing of the cluster centres
by foreground groups and clusters could explain the over-prediction of
the observed decrements by  SZ models and in particular the apparent
tendency for higher redshift clusters to have smaller SZ
decrements. However, before considering such interpretations further,
we need to clarify if this is a real observational discrepancy between
the OVRO/BIMA data and WMAP.

It will clearly be interesting to see if these WMAP results are
confirmed in the higher resolution SZ observations made using the
Planck satellite.

\section*{Acknowledgements} We thank R. Lieu, N. Afshordi, M. Bonamente, W. Frith, G. Hinshaw and J. Mittaz for useful discussions. R. Bielby acknowledges receipt of a PPARC PhD studentship.

\label{lastpage}

\end{document}